\documentclass[aps,english,twocolumn,superscriptaddress, prl]{revtex4}
\usepackage{amsmath}
\usepackage{amsthm}
\usepackage{dcolumn}   
\usepackage{bm}        
\usepackage{color}
\usepackage{amssymb}
\usepackage{graphicx}
\usepackage[dvips,letterpaper,margin=0.7 5in,bottom=0.5in]{geometry}
\usepackage{url}
\usepackage{verbatim}
\usepackage[bookmarksnumbered,pdfpagelabels=true,plainpages=false,colorlinks=true,linkcolor=blue,citecolor=red,urlcolor=blue]{hyperref}
\usepackage[rightcaption]{sidecap}
\usepackage{epsfig,graphicx,amsfonts,amsbsy}
\usepackage{dsfont}
\usepackage{amsmath,amsfonts,amsthm,amssymb,mathrsfs,bbm}
\usepackage[latin1]{inputenc}
\usepackage[english]{babel}
\usepackage[T1]{fontenc}

\newcommand{\bs}[1]{\boldsymbol{#1}}

\begin{document}
\title{\bf{Interfacial Dzyaloshinskii-Moriya interaction in Pt/CoFeB films: effect of the heavy-metal thickness}}
\author{S. Tacchi}
\affiliation{Istituto Officina dei Materiali del CNR (CNR-IOM), Unit\`a di Perugia, c/o Dipartimento di Fisica e Geologia,
Universit\`a di Perugia, I-06123 Perugia, Italy}
\author{R. E. Troncoso}
\affiliation{Departamento de F\'isica, Universidad T\'ecnica Federico Santa Mar\'ia, Avenida Espa\~na 1680, Valpara\'iso, Chile}
\author{M. Ahlberg}
\affiliation{Department of Physics, University of Gothenburg, SE-41296 Gothenburg, Sweden}
\author{ G. Gubbiotti}
\affiliation{Istituto Officina dei Materiali del CNR (CNR-IOM), Unit\`a di Perugia, c/o Dipartimento di Fisica e Geologia, Universit\`a di Perugia, I-06123 Perugia, Italy}
\author{M. Madami}
\affiliation{ Dipartimento di Fisica e Geologia, Universit\`a di Perugia, I-06123 Perugia, Italy}
\author{J. {\AA}kerman}
\affiliation{Department of Physics, University of Gothenburg,
SE-41296 Gothenburg, Sweden}
\affiliation{  Materials and Nano
Physics, Royal Institute of Technology (KTH), SE-164 40 Kista,
Sweden}
\author{ P. Landeros}
\affiliation{Departamento de F\'isica, Universidad T\'ecnica Federico Santa Mar\'ia, Avenida Espa\~na 1680, Valpara\'iso, Chile}

\begin{abstract}
We report the observation of a Pt layer thickness dependence on the induced interfacial Dzyaloshinskii-Moriya interaction in ultra-thin Pt($d_{\text{Pt}}$)/CoFeB films.
Taking advantage of the large spin-orbit coupling of the heavy metal, the interfacial Dzyaloshinskii-Moriya interaction is quantified by Brillouin light scattering measurements of the frequency non-reciprocity of spin-waves in the ferromagnet.
The magnitude of the induced Dzyaloshinskii-Moriya coupling is found to saturate to a value $0.45$ mJ$/$m${}^2$ for Pt thicknesses larger than $\sim 2$ nm.
The experimental results are explained by analytical calculations based on the 3-site indirect exchange mechanism that predicts a Dzyaloshinskii-Moriya interaction at the interface between a ferromagnetic thin layer and a heavy metal.
Our findings open up a way to control and optimize chiral effects in ferromagnetic thin films through the thickness of the heavy metal layer.
\end{abstract}
\maketitle

In the last few years the Dzyaloshinskii-Moriya interaction (DMI) \cite{Dzyaloshinskii,Moriya}, i.e. the antisymmetric exchange interaction, has been the subject of  intense research due to its capability to induce the formation of chiral spin textures, such as magnetic skyrmion lattices \cite{Bogdanov,Muhlbauer,Munzer,Yu,Yu11,Heinze,Onose,Nagaosa,Fert13} and spin spirals \cite{Uchida,Bode,Ferriani}. 
In ultrathin ferromagnetic (FM) films in contact with a nonmagnetic heavy-metal (HM), a noticeable interfacial DMI can arise due to the large spin-orbit coupling (SOC) in the presence of the broken inversion symmetry at the FM/HM interface \cite{Bode,Fert13}, leading for instance to asymmetric spin-wave dispersion \cite{Zakeri}.
Interfacial DMI in FM/HM bilayers is usually stronger than bulk DMI in non-centrosymmetric chiral magnets \cite{Iguchi,Seki}, which also has the advantage of room temperature operation using conventional magnetic materials. 
In such structures, the combination of the interfacial DMI, which stabilizes chiral N$\acute{e}$el domain walls (DW), and of the Spin-Hall effect \cite{Liu,Sinova} has been found to enable a surprisingly fast current-driven DW motion \cite{Heide,Moore,Thiaville,ChenPRL,Emori,Boulle, Brataas,Torrejon}.
It has also been observed that both the velocity and the direction of the DW motion depend on the DMI strength and can be controlled by engineering the interface between the two materials \cite{Torrejon,Ryu,Chen}. 
From a technological point of view, these structures are of great importance, due to their enormous potential for current-controlled DW motion for the development of novel memory-storage devices with high density and performance in so-called racetrack memories \cite{Parkin}. 
A deeper understanding of the interfacial DMI mechanism in such structures and a precise estimation of its magnitude, are therefore crucial for tailoring efficient spintronics devices.

\begin{figure}[ht]
\begin{center}
\includegraphics[width=3.2in]{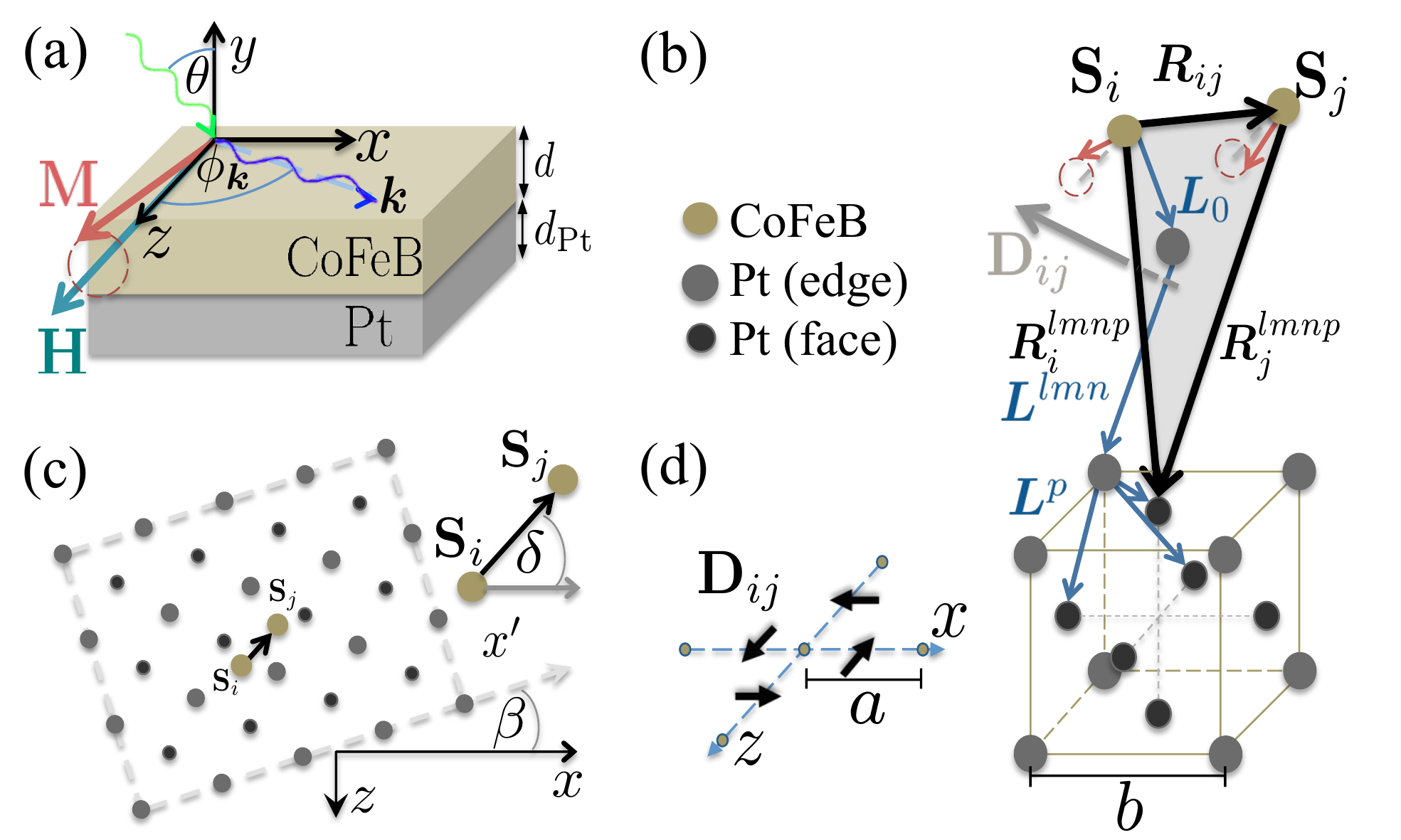}
\caption{
(a) Schematic depiction of the system under study. The magnetization ${\bf M}$ is saturated along $z$ axis by an external magnetic field ${\bf H}$. Spin-waves propagate on the $xz$-plane and are characterized by a wave-vector ${\bs k}$ making an angle $\phi_{\bs k}$ with the saturation direction. Based on the 3-site Fert-Levy model the interfacial DMI is determined under the scheme (b), where brown dots, $i$ and $j$, represent a pair of CoFeB magnetic moments interacting through a third Pt atom, illustrated by the grey dots, at position ${\bs R}_{i}^{lmnp}$ with respect to spin $i$.
(c) Illustration of a (100) plane of a fcc crystallite oriented at an angle $\beta$ with the $x$ axis, while the pair of spins is oriented at an angle $\delta$ with $x$.  
As indicated by the picture in (d), the resulting DMI vector is distributed on the $xz$-plane and perpendicular to the vector ${\bs R}_{ij}$.}
\label{fig1}
\end{center}
\end{figure}

Early measurements of the strength of the DMI were reported using spin-polarized scanning tunneling microscopy \cite{Bode}, highly resolved spin-polarized electron energy loss spectroscopy \cite{Zakeri}, and synchrotron based X-ray scattering \cite{Dmitrienko}.
More recently, Brillouin light scattering (BLS) has proven to be a powerful technique to study interfacial DMI in a variety of FM/HM systems \cite{DiPRL,Zhang,DiAPL,Cho,Nembach,Stashkevich,Belmeguenai} such as shown in Fig. \ref{fig1}.
BLS experiments on ultrathin FM/HM bilayers have shown that interfacial DMI induces a significant asymmetry in the frequency dispersion of the counter-propagating Damon-Eshbach (DE) spin-wave (SW) modes, as theoretically predicted in Refs. \cite{Kataoka,Udvardi,Costa,Cortes,Moon}, which makes direct measurements of the strength of the induced DMI possible.
Moreover, the effect of the interfacial DMI has been investigated in wedge-shaped samples \cite{Cho}, and also in structures where the thickness of the FM layer $d$ is varied \cite{Nembach,Stashkevich,Belmeguenai}, demonstrating a $1/d$ behavior of the strength of the interaction, which is direct consequence of the surface nature of such coupling. 
This phenomenology was also found through all-electrical measurements in Pt/Co/MgO samples \cite{Lee}. Interestingly, the discussion related to the role and importance of Pt thickness is devoid in all those experiments. 
More recently Yang {\it et al}. performed first principles calculations of DMI in Co/Pt where its strength is featured for specific spin configurations and up to three Co and Pt atomic layers, founding a weak contribution from Pt thickness \cite{Yang}.

In this work we study the influence of the heavy metal thickness on the interfacial DMI.
Using BLS measurements on ultrathin CoFeB films in contact with a Pt layer with variable thickness ($d_{\text{Pt}}$), we found that the strength of the interfacial DMI increases with Pt thickness, reaching a saturation value for $d_{\text{Pt}}$ larger than a few nanometers.
We are able to explain our experimental results using the $3$-site DMI introduced by Levy and Fert \cite{Levy69,Smith,Fert}, where the asymmetric exchange interaction between two neighboring FM atoms is mediated by a third non-magnetic atom, Pt in this case, having a large SOC.
Here we show that the evolution of interfacial DMI as a function of the Pt thickness, can be  understood assuming that hopping electrons can scatter with Pt sites belonging to several layers in the HM.

We studied a series of samples consisting of Si-SiO$_{2}$/Co$_{40}$Fe$_{40}$B$_{20}$(2 nm)/Pt($d_{\text{Pt}}$)/Cu(3 nm) where $d_{\text{Pt}}$ was changed in the range between 0 and 6 nm.
The samples were grown by magnetron sputtering on thermally oxidized Si substrates.
The base pressure of the chamber was $2\times 10^{-8}$~Torr, and the deposition times were calculated using calibrated growth rates.
The saturation magnetization was determined from hysteresis curves measured by a MicroMag $2900$ alternating gradient magnetometer (AGM). BLS measurements were performed focusing about $200$ mW of monochromatic light from a solid state laser operating at $\lambda=532$ nm onto the sample surface.
The back-scattered light was analyzed by a Sandercock-type ($3+3$)-pass tandem Fabry-Perot interferometer \cite{BLS}.
A bias field $H=3$ kOe was applied parallel to the surface plane, while the in-plane wave vector ${\bs k}$ was swept along the perpendicular direction (DE configuration).
Due to the photon-magnon conservation law of momentum in the scattering process, the amplitude of the in-plane wave vector is linked to the incidence angle of light $\theta$ by  $k=(4\pi/\lambda)\sin\theta$.
In our measurements $k$ was changed from 0 to $2.044\times 10^7$ rad/m.

In order to analyze the experimental results we start with the usual Hamiltonian ascribed to the interfacial DMI, ${\cal H}_{\text{DM}}=\sum_{\langle ij\rangle}{\bf D}_{ij}\cdot\left({\bf S}_i\times{\bf S}_j\right)$, which couples neighboring atomic spins ${\bf S}_i$ and ${\bf S}_j$ in the interfacial layer of the CoFeB film through a third Pt site \footnote{This type of interfacial DMI is similar to the one that should appears in non-centrosymmetric crystals of class $C_{nv}$ \cite{Cortes}, where the interaction is between two FM atomic spins with spin-orbit on the FM site \cite{Moriya}.}.
The DMI in FM/HM bilayers is usually described using a formalism developed for disordered magnetic alloys with HM impurities \cite{Fert,Prejean}. 
Here, an additional contribution to the Ruderman-Kittel-Kasuya-Yosida interaction appears, which is of Dzyaloshinskii-Moriya type and arises from the SOC of the conduction electron gas with non-magnetic impurities \cite{Fert}. 
The DM vector ${\bf D}_{ij}$ links FM spins at  sites $i$ and $j$ with a third Pt site in the HM \cite{Fert} and is perpendicular to the triangle described by the three sites. 
It is well known that the DMI becomes particularly relevant at the interface between a ultrathin FM film and a HM with strong SOC. 
This fact, together with the BLS data presented here, suggests that several Pt atoms may contribute to the strength of the interfacial DMI. 
Hence, in order to evaluate the DM vector, one have to consider the thickness and lattice structure of the HM, in such a way that the electrons can scatter with several Pt sites close to a pair of FM spins, and thus build up the effective interfacial DMI.
Then, the DM vector \cite{Fert,Crepieux} associated with ${\bf S}_i$ and ${\bf S}_j$, must include contributions from more than one Pt atom, and can be generally estimated from
\begin{align}
\label{eq1}
{\bf D}_{ij}=\frac{\lambda_{0}}{R_{ij}}\sum_{lmnp}\frac{{\bs R}_{i}^{lmnp}\cdot{\bs R}_{j}^{lmnp}}{\left(R_{i}^{lmnp}R_{j}^{lmnp}\right)^3}{\bs R}_{i}^{lmnp}\times{\bs R}_{j}^{lmnp},
\end{align}
where $\lambda_{0}$ is a factor proportional to the SOC constant and other parameters of the HM \cite{SuppMat}. 
The sum is over Pt lattice sites that are neighbors with ${\bf S}_i$ and ${\bf S}_j$.
As it is displayed in Fig. \ref{fig1}, the vectors ${\bs R}_{i}^{lmnp}={\bs L}_0+{\bs L}^{lmn}+{\bs L}^{p}$, and ${\bs R}_{j}^{lmnp}={\bs R}_{i}^{lmnp}-{\bs R}_{ij}$, join ${\bf S}_i$ and ${\bf S}_j$ to the fcc lattice sites of Pt. 
Vector ${\bs L}_0=L_{0x}\hat{x}-L_{0y}\hat{y}+L_{0z}\hat{z}$ joins FM site $i$ with the closest Pt atom labeled with $l=m=n=0$, 
while vector ${\bs L}^{lmn}=bl\hat{x}'-bm\hat{y}+bn\hat{z}'$ runs from the (000) Pt site to the neighbors cubic sites at the edges of a fcc lattice, labeled by ($lmn$), where $l,n\in[-N,N]$ and $m\in[0,N_{\text{Pt}}]$, with $N$ and $N_{\text{Pt}}$ running through a few lattice sites, corresponding to a distances of the order of the spin diffusion length. 
Vector ${\bs L}^{p}$ describes the position of the 4 Pt atoms associated to site ($lmn$): ${\bs L}_{1}=b/2(\hat{x}'-\hat{y})$, ${\bs L}_{2}=b/2(\hat{x}'+\hat{z}')$, ${\bs L}_{3}=b/2(-\hat{y}+\hat{z}')$, and ${\bs L}_{4}=0$.
CoFeB atoms are assumed in the film plane then ${\bs R}_{ij}=na(\cos\delta\hat{x}+\sin\delta\hat{z})$  
where $\delta$ is the angle between ${\bs R}_{ij}$ and the $x$ axis, while $a$ is the average separation of nearest neighbors spins.
The index $n$ is introduced to consider first, second or even third neighbors \cite{SuppMat}. 
With this model, we get for the DM vector between spins at $i$ and $i+x$, ${\bf D}_{i,j=i+x}=-D_z\hat{z}$, and between spins at $i$ and $i+z$ ${\bf D}_{i,j=i+z}=D_x\hat{x}$ [see Fig. \ref{fig1}(d)], where the $y$ component of ${\bf D}_{ij}$ cancels out \cite{SuppMat}. 

In the micromagnetic limit the DM Hamiltonian is determined \cite{SuppMat} by assuming that the magnetization does not depend on the normal coordinate, due to the ultrathin thickness of the ferromagnetic film. On this basis, the frequency dispersion of the spin waves is computed and it turns out to be separated into two contributions, $f({\bs k})=f_s({\bs k})+f_{\text{DM}}({\bs k})$, with $f_{\text{DM}}({\bs k})=\frac{\gamma {D}(d_{\text{Pt}})}{\pi M_s}|{\bs k}|\sin\phi_{\bs k}\cos\phi_M$, where ${D}(d_{\text{Pt}})\equiv\frac{S^2}{ad}|{\bf D}_{ij}|$ is the volume averaged DMI strength. 
The symmetric part, $f_s({\bs k})$, is composed by the exchange, dipolar, anisotropy, and Zeeman contributions \cite{Cortes}. Here, $\gamma=| g \mu_B/\hbar |$ is the gyromagnetic ratio, and $\phi_M$ the angle between $\bf M$ and the plane \cite{SuppMat}. Then, the frequency difference between oppositely propagating spin-waves is $\Delta f =f_{\text{DM}}({\bs k})-f_{\text{DM}}({-\bs k})$, and its dependency with the FM and HM layers reads \footnote{In the case of bulk DMI, as those encountered in B$_{20}$ crystals, a similar spin-wave theory \cite{Cortes} have shown that $\Delta f \propto k\cos\phi_{\bs k} $, which has been measured recently in chiral magnets Cu$_2$OSeO$_3$ \cite{Seki} and LiFe$_5$O$_8$ \cite{Iguchi}.}
%
\begin{align}\label{eq2}
\Delta f({\bs k},d_{\text{Pt}})=\frac{2\gamma
{D}(d_{\text{Pt}})}{\pi M_s}|{\bs k}|\sin\phi_{\bs k}\cos\phi_M,
\end{align}
where ${D}(d_{\text{Pt}})$ measures the strength of the interfacial DMI averaged over the volume of the FM film. 
According to Nembach \textit{et al.} \cite{Nembach} it is related to the DMI strength at the interface  $D_{\text{int}}={D}(d_{\text{Pt}})N_{\text {FM}}$, where $N_{\text {FM}}$ is the number of FM atomic layers. 
By measuring $\Delta f$ through BLS the DMI strength has been found in several materials, whose highest value of 2.7 mJ/m$^2$ was reported in Pt(3)/Co(0.6)/AlO$_{\rm x}$ samples \cite{Belmeguenai}.

%
\begin{figure}[ht]
\begin{center}
\includegraphics[width=4.3 in]{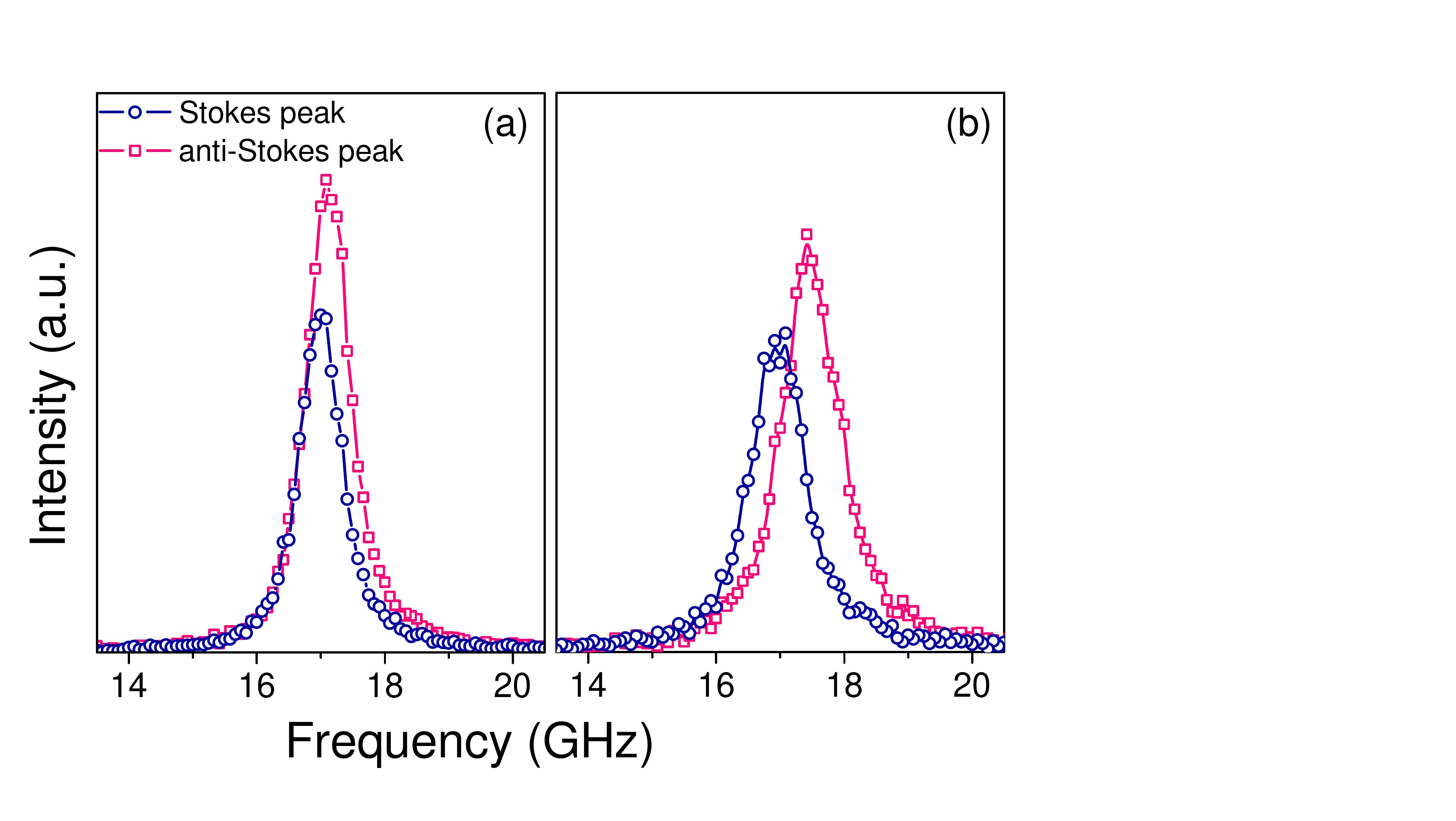}
\caption{Brillouin light scattering spectra measured at $k=1.81\times 10^7$ rad/m under a magnetic field $H=3$ kOe
for samples having a Pt thickness of {(a) $0.4$ nm and (b) 1 nm.}} \label{fig2}
\end{center}
\end{figure}

\begin{figure}[ht]
\begin{center}
\includegraphics[width=3.2in]{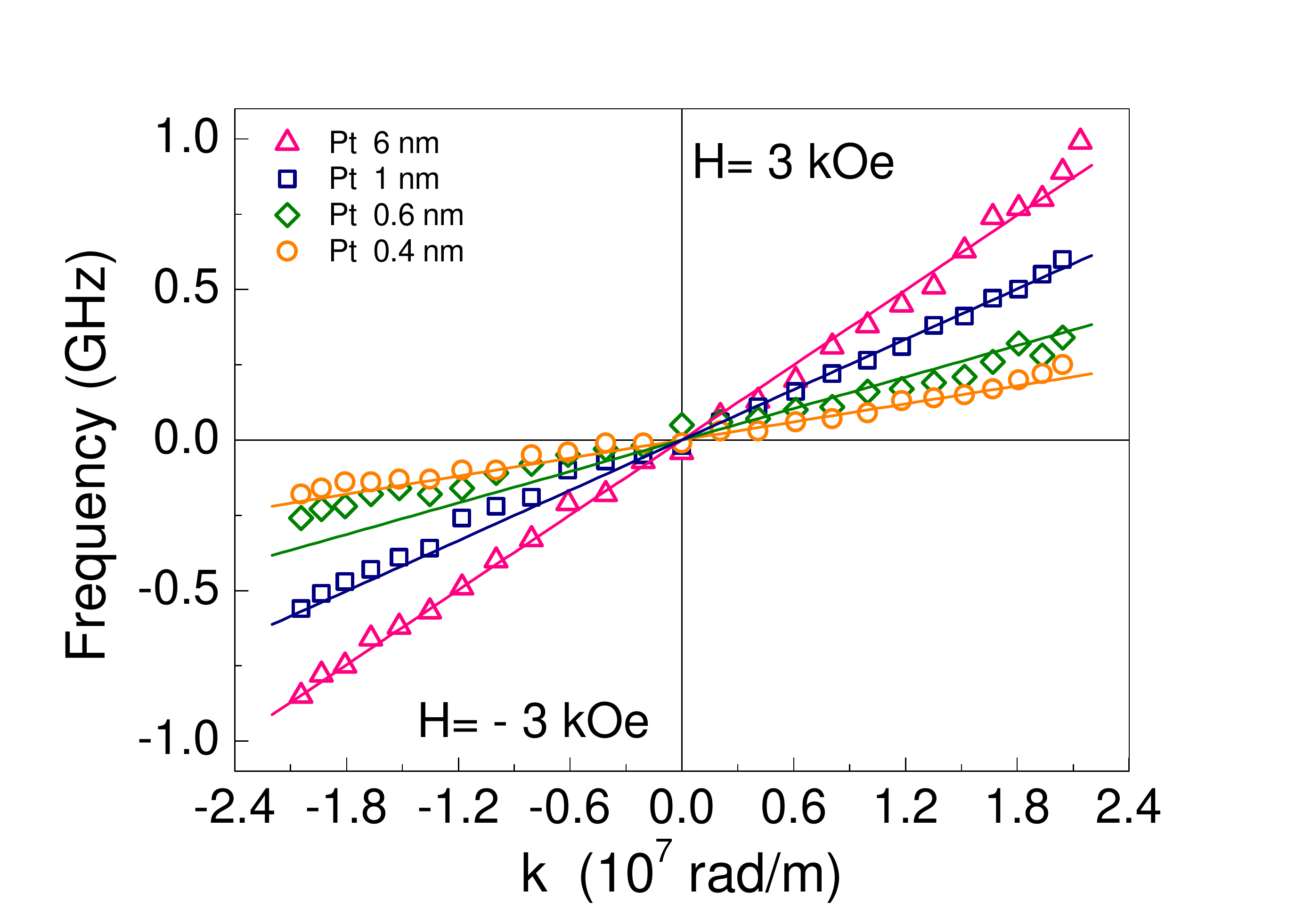}
\caption{Measured frequency shift for different Pt thickness.
The samples are under a magnetic field $H=\pm3$ kOe, with the corresponding theoretical fitting based on Eq. (\ref{eq2}).}
\label{fig3}
\end{center}
\end{figure}

\begin{figure}[ht]
\begin{center}
\includegraphics[width=3in]{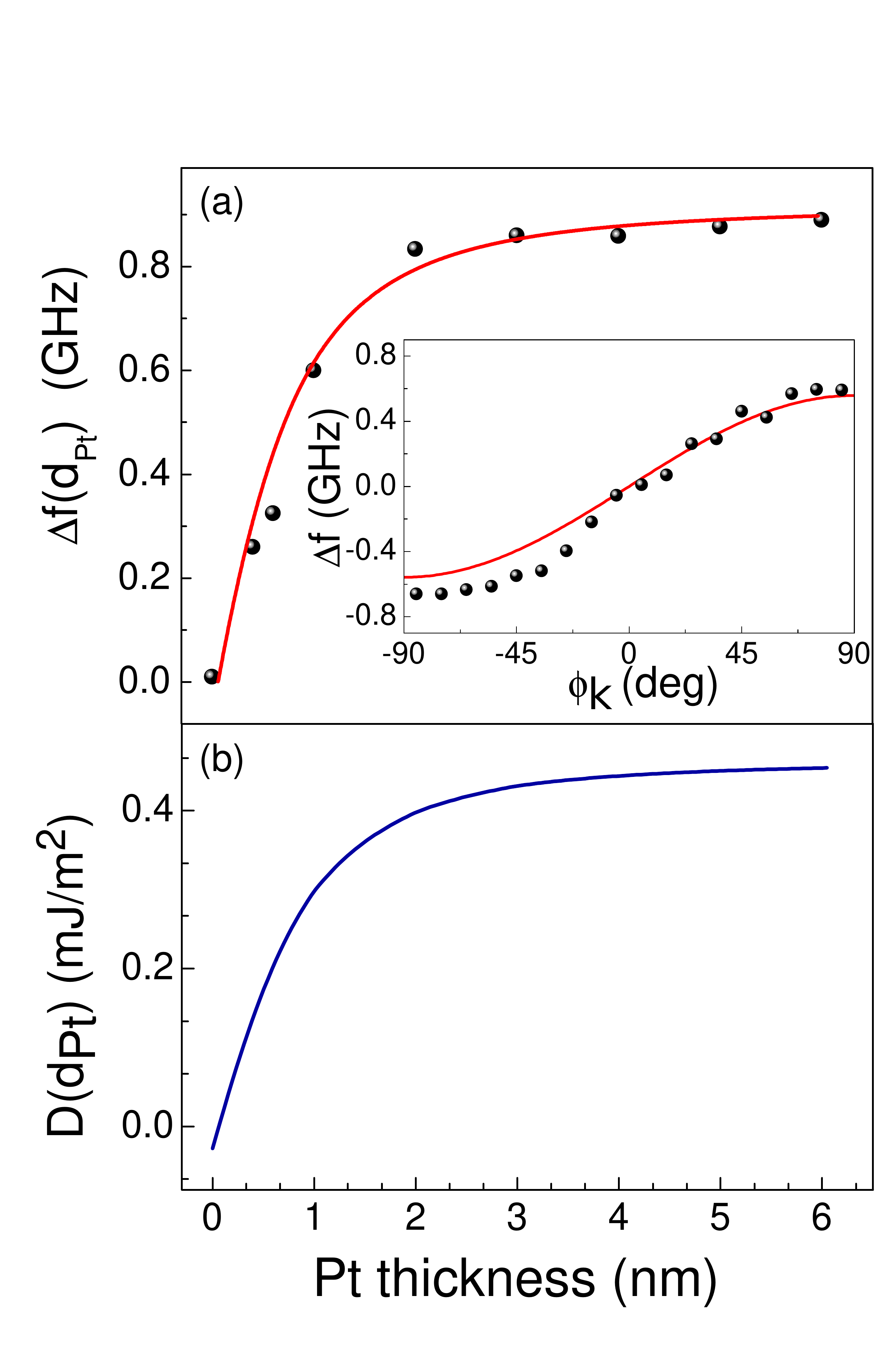}
\caption{(a) Spin-wave frequency asymmetry in a CoFeB/Pt as a function of Pt thickness at $k_{\text{max}}=2.044\times 10^7$ rad/m.
The fit (red line) of the data, based on Eq. (\ref{eq2}), is done with the parameters $a=0.25$ nm, $b=0.39$ nm and ${L}_{0y}=0.32$ nm, and $n=2.95$.
The inset shows the in-plane angular dependence of $\Delta f$, and its corresponding fit curve for a thickness $d_{\text{Pt}}=5$ nm at a wave vector $k=1.35 \times 10^7$ rad/m. (b) DMI strength as a function of $d_{\text{Pt}}$ obtained from $\Delta f(k_{\text{max}},d_{\text{Pt}})$ and for a gyromagnetic ratio $\gamma=187$ GHz/T \cite{gyrofactorCoFeB}. 
}
\label{fig4}
\end{center}
\end{figure}

Typical BLS spectra measured for samples having a Pt thickness of $0.4$ nm and $1$ nm are shown in Fig. \ref{fig2}(a) and \ref{fig2}(b), respectively. 
Due to the small sample thickness, both the Stokes and anti-Stokes peaks, corresponding to SWs propagating in opposite directions are simultaneously observed with comparable intensity. 
As it can be seen, the Stokes and anti-Stokes peaks are characterized by a frequency shift which increases with Pt thickness. 
Moreover, the frequency of both peaks interchanges on reversing the direction of the applied magnetic field, due to the reversal of the SWs propagation direction. Fig. \ref{fig3} shows the frequency difference between the Stokes and the anti-Stokes peaks measured (points) as a function of the wave vector \emph{k}.
In agreement with Eq. (\ref{eq2}) we found that the frequency asymmetry exhibits a linear dependence as a function of \textit{k}, and it becomes more pronounced when increasing the Pt thickness.
To better understand the effect of the platinum thickness, the frequency shift measured at $k_{\rm max}=2.044\times 10^7$ rad/m is reported in Fig. \ref{fig4}(a) as a function of $d_{\text{Pt}}$.
One can see that the frequency difference increases linearly with $d_{\text{Pt}}$, reaching a saturation value at about $2$ nm.
 A fit procedure \cite{SuppMat} of the experimental data to the theoretical model was performed using Eq. (\ref{eq2}).
In this analysis, we consider for amorphous CoFeB $a= 0.25$ nm \cite{Kirk}, and for the fcc lattice parameter of Pt $b=0.39$ nm \cite{Kittel}, while the atomic spacer between the CoFeB and the Pt layers, was set to the mean value between $a$ and $b$, $L_{0y}=0.32$ nm.
On the basis of the AGM measurements, we assumed $\mu_{0}M_s=1.55$ T for samples having a Pt thickness larger than 1 nm, while for samples with $d_{\text{Pt}}$ lower than 1 nm, $\mu_0 M_s$ decreases until 1.22 T.
%

A good agreement with the experiments is obtained by setting $\beta=\delta=0$, $L_{0x}=L_{0z}=0$, and the fitting parameters  $f_0\equiv\frac{2\gamma S^2\lambda_0k_{\text{max}}}{\pi M_s n^2 a^4 d } =0.03$ GHz, and the average spatial range of DMI $R_{ij}\approx na=0.74$ nm, which for $a=0.25$ nm gives $n \approx 2.95$ \cite{SuppMat}. 
The fit is shown by the continuous curves in Fig. \ref{fig3} for the linear behavior with $k$ and Fig. \ref{fig4}(a) for the thickness dependence, where the solid red curve is a linear interpolation. 
The measured angular dependence of the frequency shift, see inset of Fig. \ref{fig4}(a), was obtained for a thickness $d_{\text{Pt}}=5$ nm at a wave vector $k=1.35 \times 10^7$ rad/m, showing a clear sine like dependence \cite{Cho,Cortes,Zhang} in agreement with Eq. \ref{eq2}.
The strength of the interfacial DMI obtained from the fit is reported in Fig. \ref{fig4}(b). As it can be seen $D(d_{\text{Pt}})$ grows with the Pt thickness and reaches a saturation value of almost 0.45 mJ/m${}^{2}$ at about four Pt monolayers ($\sim 2$ nm). 
This characteristic can be traced back to the spin transport parameter, the so-called spin-diffusion length.
As we pointed out, interfacial DMI originates from the indirect exchange between FM spins and neighboring HM atoms having a large SOC. 
Such SOC in Pt is responsible for the loss of spin information carried by electrons after a characteristic spatial range given by the spin-diffusion length, which at room temperature takes values in the range of $1.2-2$ nm \cite{Liu,Obstbaum,WZhang,Isasa}.
Therefore, the spin-diffusion length spatially limits the indirect exchange mechanism that induces the DMI and accordingly, its enhancement with Pt thickness.
%

It is important to note that the average spatial range of the interfacial DM coupling $na$, is almost three times larger than the average interatomic distance $a\approx 0.25$ nm \cite{Kirk}.
This can be understood taking into account the fact that the interfacial DMI is an indirect  exchange coupling between FM atomic spins mediated by conduction electrons that populates the structure.
As a consequence two interacting spins may communicate via DMI even if they are next nearest-neighbors.
This can be verified by noting that in the present case of a saturated in-plane magnetization, the only important DMI is between the dynamic magnetization components.
Indeed, a simple evaluation of the DMI energy between spins $i$ and $j$ leads to $h_{\text{DM}}(n)=-|{\bf D}_{ij}|s_is_{j}\sin\left(kna\right)$ where $s_i$ is the dynamical component of the $i$-th spin.

Therefore, for typical wave vectors ($\sim 10^7$ rad$/$m) the DM coupling turns out to be relevant not only for first nearest neighbor precessing spins but it covers a couple of atomic sites of the CoFeB.
Similar considerations that involve the interaction of pair of spins beyond first neighbors has also been mentioned in Ref. \cite{Yang}, where the authors argued that second neighbors do not significantly contribute to the effective DMI, while direct exchange was considered up to seventh neighbors \footnote{ As it can be seen, $D$ goes rapidly to zero when $n>3$, indicating that interfacial DMI is significant up to the third neighbor spins, while its role becomes negligible for farther away spins.}.

In summary, non-reciprocity of the spin-wave spectra in Pt($d_{\text{Pt}}$)/CoFeB ultrathin films was studied by Brillouin spectroscopy for different Pt thicknesses.
The BLS spectra of Stokes and anti-Stokes peaks establishes a linear relation between the asymmetry in the SWs frequency and the wave vector.
We observed, and theoretically demonstrated by virtue of the 3-site indirect exchange mechanism \cite{Fert}, an increasing interfacial DMI as the Pt thickness increases.
We propose that the mechanism behind the observed DMI enhancement with $d_{\text{Pt}}$, consists of cumulative electron hopping between the atomic spins at the interface and the non-magnetic atoms in the heavy metal.
Nevertheless, for a given thickness of the CoFeB layer, the DMI magnitude does not exceed the saturation value $0.45$ mJ$/$m${}^2$ for Pt thicknesses larger than the spin-diffusion length $\sim 2$ nm \cite{Isasa}.
Thickness-dependent DMI studies will offer a great prospect in the fields of spintronics and magnonics, in order to induce and spatially control chiral effects in magnetic materials.

We acknowledge financial support from the G\"{o}ran Gustafsson Foundation, the Swedish Research Council (VR), Energimyndigheten (STEM), the Knut and Alice Wallenberg Foundation (KAW), the Carl Tryggers Foundation (CTS), and the Swedish Foundation for Strategic Research (SSF). 
This work was also supported by the European Research Council (ERC) under the European Community's Seventh Framework Programme (FP/2007-2013)/ERC Grant 307144 "MUSTANG", by FONDECYT (Chile) grants 3150372 and 1161403, and Centers of excellence with Basal/CONICYT financing, grant FB0807, CEDENNA.

\end{document}